\documentclass[usenatbib]{mnras}

\usepackage{newtxtext,newtxmath}
\usepackage[T1]{fontenc}
\usepackage{ae,aecompl}
\usepackage{xcolor}
\definecolor{ao(english)}{rgb}{0.0,0.5,0.0}

\usepackage[utf8]{inputenc}
\usepackage[T1]{fontenc}

\usepackage{graphicx}	
\usepackage{amsmath}	
\usepackage{systeme}    
\usepackage{subcaption} 
\usepackage{breqn}      

\newcommand{\degree}{°}
\newcommand{\revision}[1]{#1}

\title[A search for magnetic $\delta$\,Scuti stars]{A search for magnetic $\delta$\,Scuti stars in {\it Kepler} hybrid candidates}

\author[K. Thomson-Paressant et al.]{
K. Thomson-Paressant,$^{1}$\thanks{E-mail: keegan.thomson-paressant@obspm.fr}
C. Neiner,$^{1}$
P. Lampens,$^{2}$
J. Labadie-Bartz,$^{1}$
R. Monier,$^{1}$
\newauthor{
P. Mathias,$^{3}$
A. Tkachenko$^{4}$}
\\
$^{1}$LESIA, Paris Observatory, PSL University, CNRS, Sorbonne University, Université Paris Cit\'e, 5 place Jules Janssen, 92195  Meudon, France\\
$^{2}$Koninklijke Sterrenwacht van Belgi\"e, Ringlaan 3, 1180 Brussel, Belgium\\
$^{3}$IRAP, Université de Toulouse, CNRS, UPS, CNES, 57 avenue d’Azereix, 65000 Tarbes, France\\
$^{4}$Institute of Astronomy, KU Leuven, Celestijnenlaan 200D, 3001  Leuven, Belgium
}

\date{Accepted XXX. Received YYY; in original form ZZZ}

\pubyear{2020}

\begin{document}
\label{firstpage}
\pagerange{\pageref{firstpage}--\pageref{lastpage}}
\maketitle

\begin{abstract} 
Numerous candidate hybrid stars of type $\delta$\,Scuti  - $\gamma$\,Doradus have been identified with the {\it Kepler} satellite. However, many of them lie outside the theoretically expected instability strip for hybrid pulsation, where $\delta$\,Sct and $\gamma$\,Dor pulsations can be simultaneously excited. We postulate that some of these pulsating stars may not be genuine hybrid pulsators but rather magnetic $\delta$\,Sct stars, for which the rotational modulation from spots on the surface associated to the magnetic field produces frequencies in the same domain as $\gamma$\,Dor pulsations. We search for the presence of a magnetic field in a small sample of selected hybrid $\delta$\,Sct - $\gamma$\,Dor stars using spectropolarimetry. At the time of observations, the only $\delta$\,Sct star known to have a magnetic field was HD\,188774 with a field strength of a few hundred Gauss. Our observations were thus tailored to detect fields of this typical strength. We find no magnetic field in the hybrid candidate stars we observed. However, two of the three other magnetic $\delta$\,Sct stars discovered since these observations have much weaker fields than HD\,188774, and are of dynamo origin rather than fossil fields. It is likely that our observations are not sensitive enough to detect such dynamo magnetic fields in the cooler stars of our sample if they are present. This work nevertheless provides reliable upper limits on possible fossil fields in the hotter stars, pointing towards typically weaker fields in $\delta$\,Sct stars than in OBA stars in general. 
\end{abstract}




\section{Introduction}

$\delta$\,Scuti ($\delta$\,Sct) and $\gamma$\,Doradus ($\gamma$\,Dor) stars have masses between 2.5 and 1.4 M$\odot$ and span the A-F spectral range. They pulsate primarily in high-frequency \emph{p-}modes ($\gtrsim 5$ d$^{-1}$) and low-frequency \emph{g-}modes 
 ($\lesssim 5$ d$^{-1}$) respectively, and hybrids of these two types also exist since their instability strips overlap. Such hybrid pulsators are very interesting targets for asteroseismic studies as their pulsations can constrain different layers inside the star \citep{kurtz2022}.

Prior to the first high-precision photometric satellite observations, such as those from the Convection, Rotation and planetary Transits satellite \citep[{\it CoRoT}; ][]{auvergne2009}, the {\it Kepler} satellite \citep{borucki2010}, and the Transiting Exoplanet Survey Satellite \citep[{\it TESS}; ][]{Ricker2015}, very few hybrid pulsators were known \citep[e.g.][]{henry2005}. Thanks to the space-based large scale surveys and the precision of their results, \revision{hundreds of potential hybrid candidates have since been identified}. In particular, data collected by the {\it Kepler} mission \citep{grigahcene2010,uytterhoeven2011} suggest that the number of hybrid candidates is much higher than expected. If these candidates were indeed confirmed to be true hybrids, current theory would need to be seriously revised \citep[see][]{grigahcene2010,balona2015a}, and thus it is very important to assess whether they are indeed genuine hybrid pulsators.

If we assume these candidates are \emph{not} $\delta$\,Sct-$\gamma$\,Dor hybrids, two alternative explanations for the presence of low frequency variability within these stars can be put forward:
\begin{itemize}
\item either the $\delta$\,Sct star is a member of a binary (or multiple star) system, in which case the low order frequencies could originate from the orbital period of an eclipsing or ellipsoidal system. Additionally, the $\delta$\,Sct star could host tidally-excited \emph{g-}modes pulsations or could be distorted by tidal interactions with its companion(s). There is also the possibility of the companion being an unresolved $\gamma$\,Dor variable while the $\delta$\,Sct frequencies would appear to originate from the target star.
\item or, due to the presence of a magnetic field, the $\delta$\,Sct star could display some surface inhomogeneity. In this case the low frequencies would be attributed to rotational modulation rather than \emph{g-}modes. 
\end{itemize}

\begin{table*}
\begin{tabular}{lccccccl}
\hline
Target & Mass & T$_{\rm eff}$ & $v$sin$i$ & $B_{l}$ & $B_{\rm pol}$ & Spectral Type & Magnetic characterisation \\
 & (M$_\odot$) & () & (km\,s$^{-1}$) & (G) & (G) & \\
\hline\hline
HD\,188774 & $2.61 \pm 0.1$ & $7600 \pm 30$ & $52 \pm 1.5$ & [23 : 76] & $\sim$ 250 & A7.5IV-III & Dipole fossil field \\
$\rho$ Pup & $1.85 \pm 0.2$ & $6675 \pm 175$ & $8 \pm 0.4$ & [-0.29 : -0.05] & $\sim$1 & F2IIm & Ultra-weak fossil field \\
$\beta$\,Cas & $1.91 \pm 0.02$ & $7080 \pm 20$ & $70 \pm 1$ & [-6 : 4] & $\sim$20 & F2III & Dynamo field \\
HD\,41641 & $2.3\pm 0.6$ & $7200\pm80$ & $30 \pm 2$ & [-178 : 182] & 1055 & A5III & Complex fossil field \\
\hline\hline
\end{tabular}
\caption{Table of information regarding the four confirmed magnetic $\delta$\,Sct stars discovered to date, with values being retrieved from their respective papers \citep{lampens2013,neiner2017,zwintz2020,thomson2020}. Columns 2, 3 and 4 detail the stellar parameters namely stellar mass, effective temperature, and rotational velocity respectively. Column 5 provides the range of observed longitudinal field values, while column 6 gives an estimate of the polar field strength assuming a dipolar field.}
\label{tab:delta_scuti}
\end{table*}

Magnetism invariably leads to a photometric signal at the stellar rotation period due to surface inhomogeneities, both for large-scale fossil fields  \citep[e.g.][]{daviduraz2019} and for smaller-scale dynamo fields \citep[e.g.][]{2013ApJ...777..153A, 2021ApJS..255...17S}. Thus, rotational variable stars are prime candidates for magnetic measurements.  However, rotation periods may be long (up to years or decades in extreme cases), and/or photometric amplitudes may be low so that not all magnetic stars will exhibit a detectable photometric rotational signal over a given time baseline.

In a study of the candidate hybrid star HD\,188774, \citet{lampens2012} determined that the low frequencies in the variations of the light and radial velocity were related to each other: one harmonic and its parent frequency, but the phased curves were not corresponding to those of a binary system. As a result, they excluded binarity as a source of the low frequencies present. Therefore, \citet{neiner2015} tested the latter of the two above hypotheses for the same star and found that this star possesses a weak magnetic field. The rotation period inferred from subsequent magnetic field measurements (Neiner et al., in prep.) corresponds to the lowest frequency which, together with the first harmonic, was initially attributed to $\gamma$\,Dor pulsations  \citep{uytterhoeven2011}. Thus, HD\,188774 is not a genuine hybrid pulsator but a magnetic $\delta$\,Sct star. The relatively simple observed Zeeman signatures point to a fossil origin (as for OB and Ap stars) of the $258\pm 63$ G field, rather than to a dynamo field. This is also consistent with the long-term stability of the spots at the surface of HD\,188774, verified in the {\it Kepler} lightcurve \citep{lampens2013}. Since then however, a few other magnetic $\delta$\,Sct stars have been discovered, with a more complex fossil field \citep[HD\,41641,][]{thomson2020}, a dynamo field \citep[$\beta$\,Cas,][]{zwintz2020}, or an ultra-weak field \citep[$\rho$\,Pup,][]{neiner2017}. Magnetism in $\delta$\,Sct stars thus seems rather diverse. The properties of these four confirmed magnetic $\delta$\,Sct stars are summarised in Table~\ref{tab:delta_scuti}. Note that all four are somewhat evolved stars.

There have been \revision{several} other claims of magnetism found in $\delta$\,Sct stars. \revision{For example, HD\,21190 was found to be a magnetic $\delta$\,Sct star by \cite{kurtz2008} and \cite{hubrig2016} but, while the $\delta$\,Sct status was confirmed by \cite{barac2022}, \cite{bagnulo2012} showed that the magnetic detection was spurious. HD\,35929 is another potential candidate found by \cite{alecian2013}, but their magnetic detection was only tentative and this star is a pre-main sequence Herbig star.} Other magnetic $\delta$\,Sct targets suggested in the literature \revision{are assumed to be magnetic stars due to their Ap nature or spectral properties but had no spectropolarimetric measurements performed yet \citep[e.g.][]{murphy2020} or have a detected magnetic field but are not confirmed $\delta$\,Sct stars so far \citep[e.g.][]{hubrig2023}. These stars are very good magnetic $\delta$\,Sct candidates but since they are not confirmed yet they are not included in Table~\ref{tab:delta_scuti}. Indeed, for a few such candidates a spectropolarimetric study was performed and did not detect the presence of a magnetic field \citep[e.g. $\beta$\,Pic,][]{zwintz2019}. It is thus necessary to remain conservative before 
listing stars as bona fide magnetic $\delta$\,Sct stars.}

\section{Target selection}
\subsection{Sample}

Following the detection of a magnetic field in HD\,188774 \citep{neiner2015}, we decided to search for magnetic fields in several other $\delta$\,Sct - $\gamma$\,Dor hybrid candidate stars from the {\it Kepler} mission.
An initial sample of about 50 bright A/F-type hybrid candidates were proposed by \citet{uytterhoeven2011}, using complementary, follow-up spectroscopic data collected with the HERMES spectrograph \citep{raskin2011} attached to the Mercator telescope \citep{lampens2018}. We then selected targets from this sample that either are hotter than the [6900\,-\,7400]~K temperature range where hybrid pulsations are predicted by theoretical models \citep{dupret2005} -- this concerns BD+41\degree3389, HD\,175841, HD\,175939, HD\,181206, HD\,181569, HD\,183280, and HD\,226284 --, or show apparently complex low frequency variability which could include signals related to rotation -- e.g. BD+42\degree3370 and HD\,185115. In addition, BD+40\degree3786 and two stars without $\delta$\,Sct pulsations -- HD\,187254 and HD\,178874 (see Fig.~\ref{fig:periodograms}) --, were chosen based on a preliminary classification as candidates for stellar activity/rotational modulation \citep[][cf. their figs.~7~b and c]{uytterhoeven2011}. In this way, we pre-selected a set of 12 targets of {\it Kepler} magnitude (Kp) between 6.9 and 9.9 mag.

The spectral types of the selected targets range from A5 to F3 while the $v$sin$i$ values go from 15 to 240 km\,s$^{-1}$. The Lomb-Scargle periodograms of various candidates display low-frequency peaks with the ratio (3:)2:1, which might indicate that rotation is a key player. This was indeed the explanation for the two most dominant low frequencies detected in the light and radial velocity curves of the first detected magnetic $\delta$\,Sct star, HD\,188774 \citep{neiner2015}. 

While there was no direct evidence of binarity based on the multi-epoch HERMES spectra for all our targets, this was later on invalidated for three of them. Indeed, HD\,175939, HD\,185115, and BD+42\degree3370 were all previously classified as `P+VAR' with `VAR' meaning `low-amplitude RV variability for a yet unknown reason' in \cite{lampens2018}, but they meanwhile turned out to be long-period spectroscopic binaries \citep{lampens2021}.

\subsection{{\it Kepler} frequency analysis} \label{sec:Kepler_analysis}

The \textsc{Period04} package \citep{Lenz2005} was used to generate periodograms from the {\it Kepler} photometry to measure periodic signals in the standard way, using photometric amplitudes rather than unit-less power.
There are generally three types of signals of interest to this study: high-frequency $\delta$\,Sct pulsation \citep[usually $\gtrsim$~5~d$^{-1}$, but can also be found at lower frequencies,][]{2015MNRAS.452.3073B}, low-frequency $\gamma$\,Dor pulsation (usually $\lesssim$~5~d$^{-1}$), and rotational modulation. Rotational frequencies of our targets must be below $\sim$4~d$^{-1}$, which is the critical rotation frequency for an A/F type main sequence star (the critical rotation frequency only decreases for more evolved stars). Besides the expected frequency range of these signals, other patterns in the periodograms can aid in the interpretation of the detected signals. Rotational modulation is generally non-sinusoidal in its photometric signature, and thus harmonics of the rotational frequency are expected \citep[e.g.][]{buysschaert2018}. $\gamma$\,Dor pulsation is usually characterised by groups of closely spaced frequencies \citep{2011MNRAS.415.3531B, 2019MNRAS.490.4040A}, and $\delta$\,Sct pulsators exhibit multiple pulsation modes which may span a wide range of frequencies or may be more clustered in groups \citep[but usually at significantly higher frequencies than the $\gamma$\,Dor modes;][]{2011MNRAS.417..591B,2019MNRAS.490.4040A}. 

All of our targets were observed by {\it Kepler} in long cadence (LC) mode (30-minute cadence), and all but one (KIC 9775454 = HD\,185115) were also observed in short cadence (SC) mode (1-minute cadence). Both the LC and SC data have benefits and limitations. The LC photometry tends to have (often significantly) longer observational baselines and thus provides a higher frequency resolution.
However, with a Nyquist frequency of 24 d$^{-1}$, there are challenges in measuring high frequency signals (including, in principle, $\delta$\,Sct pulsation). A further complication is that any signals above the Nyquist frequency will be reflected to lower frequencies, potentially causing aliased peaks even in the low frequency regime \citep{2012MNRAS.422..665M}. 
SC data, with a Nyquist frequency of 720 d$^{-1}$ excels at detecting high frequency signals and does not suffer from this `alias reflection' problem, but with shorter time baselines (typically 10 to 30 d for our sample) the frequency resolution is poor. We considered both the LC and SC data for our sample (barring HD\,185115 with only LC data) to i) obtain high frequency resolution, ii) ensure that the (especially lower frequency) peaks are genuine and not aliases, and iii) probe out to higher frequencies.
For stars where the periodograms generated from the SC data are not shown, they do not add any information not already communicated by the LC data (due to a higher noise floor and lower frequency resolution in the SC data and a lack of significant signals beyond 24 d$^{-1}$).

\begin{figure*}
    \centering
    \includegraphics[width=2\columnwidth]{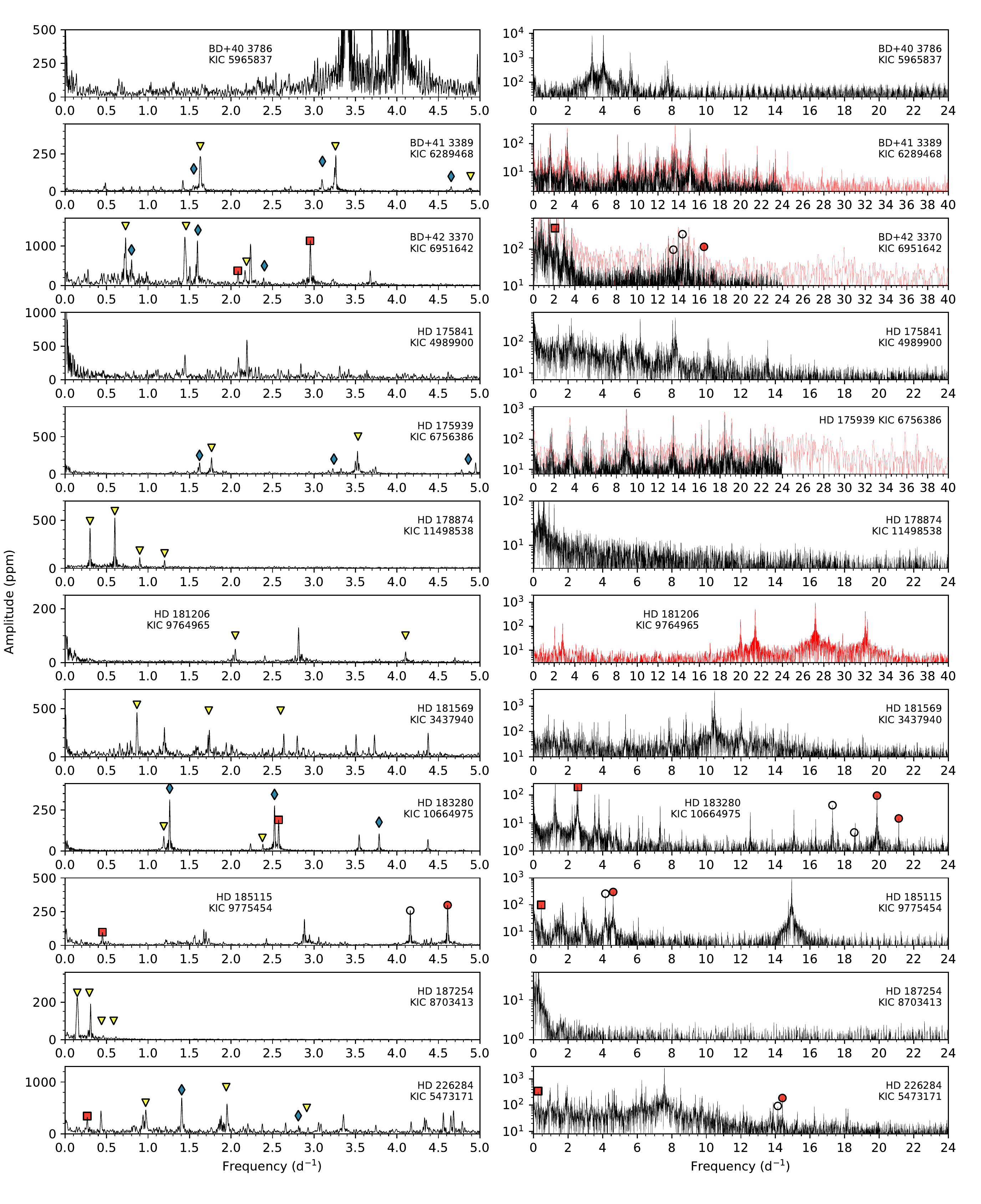} 
    \caption{Frequency spectra (periodograms) from {\it Kepler} photometry for the sample, from Cycle 6 (black), and in some cases from short cadence data (red). The left column shows the low frequency regime ($<$5 d$^{-1}$), and the right column goes out to 24 d$^{-1}$, or 40 d$^{-1}$ for the four stars with frequencies above 24 d$^{-1}$ (detected in the short cadence data, which has a higher noise floor due to shorter observing baselines). A linear scale is used in the left column, and a log scale on the right (to make low amplitude signals more visible). Triangle and diamond symbols mark harmonic series. Red squares mark low frequencies equal to the difference between two higher frequency signals (indicated by filled and open circles). For HD\,183280, both pairs of filled and open circles correspond to the same difference frequency.}
    \label{fig:periodograms}
\end{figure*}

The main results of our frequency analysis are shown in Fig.~\ref{fig:periodograms}. The periodograms for each star are separated into two regimes, up to 5 d$^{-1}$ (where rotation and $\gamma$\,Dor pulsation are most prominent), and up to 24 d$^{-1}$ (or to 40 d$^{-1}$ for the four stars with higher frequencies in the SC data). 
For all the stars in the sample, none of the low frequencies ($<$~5~d$^{-1}$) are `Nyquist reflections' of higher frequencies. However, we caution that in some cases there are relatively low amplitude signals higher than 24 d$^{-1}$ so that any quantitative analysis of the $\delta$\,Sct pulsations should consider the SC data.
Candidate rotational frequencies (identified as such by the presence of one or more harmonics) are seen in all but three of the targets.
In some cases a second harmonic series is detected, so that the candidate rotational frequency of the star is ambiguous with photometry alone. However, for this study the important characteristic is the (likely) presence of rotational modulation even if the `true' stellar rotational frequency cannot be determined with the available data. 

The three stars without any clear harmonics in the low-frequency regime (HD\,175841, BD+40\degree3786, HD\,185115) probably owe their low frequency signals to pulsation. In the case of BD+40\degree3786, the non-detection of rotational modulation in our study contradicts the indications shown by \citet{uytterhoeven2011}. It and the two other previously mentioned stars were nevertheless kept in the sample for confirmation and comparison. $\gamma$\,Dor pulsation typically follows a pattern where signals are found in densely-packed `frequency groups', the strongest located near the stellar rotational frequency, the next strongest at twice the stellar rotational frequency, and so on \citep{2018MNRAS.477.2183S}. HD\,175841 and BD\,+40\degree3786, however, do not follow this pattern and thus their frequency spectra are more ambiguous. In addition, HD\,185115 is a long-period binary and this system could contain a $\delta$\,Sct component and a $\gamma$\,Dor component.

Even among those with candidate rotational modulation, there are often additional signals in the low-frequency regime which are unrelated to the harmonic structures and could better be explained by pulsation. Stars with low-frequency signals that do not correspond to rotation may be genuine hybrid pulsators, could be cases where $\delta$\,Sct pulsations extend into the lower frequency regime, or an undetected binary with a $\delta$\,Sct component and a $\gamma$\,Dor component.

In some $\delta$\,Sct stars, low- and high-frequency pulsation modes can couple \citep[e.g.][]{2012ApJ...759...62B}. While a comprehensive analysis of all detected frequencies in our sample is beyond the scope of this work, there are four stars where such a coupling seems evident.
These are indicated in Fig.~\ref{fig:periodograms}, where the spacing between the two higher-frequency signals marked with open and filled circles is equal to the low-frequency signal marked by a filled square.
HD\,183280 has two pairs of high frequency signals with the same spacing.
BD+42\degree3370 has two low frequency signals which both seem to couple to the same high frequency signal. For a detailed analysis of its pulsations, we refer to the study by \citet{samadi2022}. The authors discovered many multiplets of rotationally split \emph{g-} and \emph{p-}modes as well as the candidate rotation frequency with two harmonics in the low region of the Fourier spectrum.
Such coupling may be of asteroseismic interest, especially if rapid rotation is somehow related  \citep[as hypothesized in][]{2012ApJ...759...62B}

The candidate rotational signals identified in the {\it Kepler} data of our sample are similar to those found in the two known magnetic $\delta$\,Sct pulsators with relatively strong fossil fields, listed in Table~\ref{tab:delta_scuti}. 
In HD\,41641, which hosts a complex fossil field, there is a rotational signal at 2.8077~d (0.3562~d$^{-1}$) with two clear harmonics seen in the {\it TESS} space photometry \citep[extracted and analysed by us, see Appendix~\ref{sec:ap1}, and in agreement with {\it CoRoT} observations by][]{2016A&A...588A..71E}. 
HD\,188774, with a dipole fossil field, also has a rotational signal (with harmonics) in {\it Kepler} data with P = 2.90711~d (0.3440~d$^{-1}$).
In $\beta$\,Cas, with a weak dynamo field, {\it TESS} could not detect a rotational frequency \citep{zwintz2020}, perhaps due to a low photometric amplitude, even though with $P_{\rm rot} \approx$ 
0.868 d, about 57 rotation cycles were covered by the analysed {\it TESS} data. {\it TESS}, however, has lower photometric precision compared to {\it Kepler} (and {\it Kepler} did not observe this star). 
Another potential rotation period value has recently been put forward for $\beta$\,Cas at 1.518 d (V. Antoci, private communication). This period is visible with very low amplitude in the {\it TESS} data, as well as a potential first harmonic. 
Rotational modulation is not seen in the {\it TESS} photometry (see Appendix~\ref{sec:ap1}) of $\rho$\,Pup (with an ultra-weak fossil field). $\rho$ Pup is a very slow rotator \citep[$v$sin$i$ = 8 $\pm$ 0.4 km\,s$^{-1}$,][]{2012A&A...542A.116A}, which makes rotation more difficult to detect in the relatively short duration {\it TESS} lightcurves.

\section{Spectropolarimetric observations}
The targets were observed with the Echelle SpectroPolarimetric Device for the Observation of Stars \citep[ESPaDOnS,][]{espadons2006}, operating on the Canada France Hawaii Telescope (CFHT) at the Mauna Kea Observatory, Hawaii. The observations were taken over a period of three months, between 2016 April 17 and 2016 June 24. The list of targets and their respective observations are available in Table~\ref{tab:full_table}. The exposure time used for each target was defined by considering the stellar properties (in particular temperature and $v$sin$i$) and assuming that the field possibly present in the targets is of the same order of magnitude ($\sim$250 G) as the one already detected in HD\,188774 (which was the only magnetic $\delta$\,Sct known at the time of observations). In practice this threshold also depends on weather conditions during observations, as the signal-to-noise ratio (SNR) we can achieve for a given observation can vary. Successive Stokes sequences were taken for most targets, in order to reach the overall SNR required for that theoretical magnetic field threshold (see Table \ref{tab:full_table}).

A near identical process as the one described in \cite{thomson2020} was followed for these targets, which we will summarise briefly over the next few paragraphs. Data taken from the ESPaDOnS instrument was reduced with the \textsc{LibreEsprit} \citep{donati1997} and \textsc{Upena} \citep{martioli2011} pipelines. We then generated three spectra: Stokes I, the flux spectrum; Stokes V, circular polarisation made by combining four sub-exposures; and a null polarisation spectrum, labelled N, generated by combining the sub-exposures destructively and used to check for pollution in the spectra. The one difference, compared to \cite{thomson2020}, is that prior to running the Least Squares Deconvolution (LSD) method \citep{donati1997}, as described below, we performed continuum normalisation using \textsc{SpeNT} \citep{martin2017} for all the spectra of each target, as we noticed the automated normalisation done by \textsc{LibreEsprit} did a poor job of fitting the blue region of the spectra. Additionally, we calculated synthetic spectra with the \textsc{SYNSPEC49} code \citep{hubeny1992} - combined with the \textsc{ATLAS9} \citep{kurucz1992} and \textsc{ATLAS12} \citep{kurucz2005,kurucz2013} codes for atmospheric and chemical abundance modelling - and utilised them to ensure what we were fitting was indeed the continuum and not a "pseudo-continuum" due to the blending of many lines in the blue part of the spectrum. Synthetic spectra were generated using the stellar parameters of the stars (notably T$_{\text{eff}}$, log g, and $v$sin$i$, cf. Tables~\ref{tab:full_table} \& \ref{tab:upperlims}) as a closest approximation, and provided both normalised and unnormalised versions of the spectrum to use as reference.

The spectral line profiles and Stokes V profiles were averaged together using the LSD algorithm, which works by combining the available lines in each spectrum, using their respective wavelength, line depth, and Land\'e factors as weights. Combining the lines in this way increases the SNR as compared to a single line, and thus increases the averaged line sensitivity to the presence of a magnetic field.

\begin{table*}
\begin{tabular}{c@{\,}ccccc@{\,}ccccc}
\hline
Target & KIC & Date & Mid-HJD & T$_{\text{exp}}$ & Mean $\lambda$ & Mean Landé & SNR & FAP & $B_l \pm \sigma B$ & $N_l \pm \sigma N$ \\
 & & & +2457000 & (s) & (nm) & & & & (G) & (G) \\
\hline
\hline 
BD\,+40\degree3786 & 5965837 & 17-Apr-16 & 496.1275 & 1x4x637  & 534.2117 & 1.190 & 1898 & ND & $ 10.7 \pm 5.4 $ & $ -3.6 \pm 5.4 $ \\
          \hline
BD\,+41\degree3389 & 6289468 & 15-May-16 & 524.0308 & 2x4x1318 & 519.4100 & 1.196 & 6715 & ND & $ 0.75 \pm 116.95 $ & $ 131.65 \pm 117.05 $ \\
                   &         & 17-May-16 & 525.9622 & 2x4x1318 & 520.5958 & 1.196 & 6857 & ND & $ 23.75 \pm 125.1 $ & $ -43.85 \pm 125.2 $  \\
          \hline
BD\,+42\degree3370 & 6951642 & 19-Apr-16 & 498.0733 & 2x4x1498 & 531.7979 & 1.191 & 7106 & ND & $ -23.25 \pm 57.1 $ & $ 52.4 \pm 57 $ \\
                   &         & 22-Apr-16 & 501.1040 & 1x4x1498 & 532.0277 & 1.191 & 5010 & ND & $ -9.8 \pm 59.4 $   & $ -61.3 \pm 59.6 $ \\
                   &         & 14-May-16 & 523.0732 & 1x4x1498 & 529.9560 & 1.191 & 4844 & ND & $ -77.6 \pm 65.2 $  & $ -115.7 \pm 65.3 $ \\
          \hline
HD\,175841         & 4989900 & 17-Apr-16 & 496.0864 & 1x4x987  & 519.3916 & 1.202 & 4885 & ND & $ -42.3 \pm 74.6 $ & $ -74.8 \pm 74.5 $ \\
                   &         & 19-Apr-16 & 497.9762 & 1x4x987  & 520.4548 & 1.202 & 3801 & ND & $ 21.7 \pm 70.3 $ & $ 70.4 \pm 70.6 $ \\
          \hline
HD\,175939        & 6756386 & 19-Apr-16 & 498.9897 & 1x4x1231 & 541.8455 & 1.196 & 5798 & ND & $-77.9 \pm 89.6$ & $-68.0 \pm 89.8$ \\
                  &         & 20-Apr-16 & 500.0299 & 2x4x1231 & 540.0611 & 1.196 & 8209 & ND & $28.3 \pm 83.8$ & $68.3 \pm 83.7$ \\
                  &         & 21-Apr-16 & 500.9817 & 1x4x1231 & 542.1242 & 1.196 & 5768 & ND & $82.4 \pm 94.8$ & $60.9 \pm 94.4$ \\
          \hline
HD\,178874        & 11498538 & 18-May-16 & 527.0393 & 1x4x698 & 532.8074 & 1.192 & 2617 & ND & $ 8.4 \pm 13.5 $ & $ 7.2 \pm 13.6 $ \\
                  &          & 19-May-16 & 527.9048 & 1x4x698 & 537.8895 & 1.192 & 2678 & ND & $ 0.7 \pm 10.1 $ & $ -8.2 \pm 10.0 $ \\
                  &          & 10-Aug-16 & 611.4466 & 4x4x593 & 523.0904 & 1.194 & 3468 & ND & $ -2.43 \pm 16.4 $ & $ -0.025 \pm 16.4 $ \\
          \hline
HD\,181206        & 9764965 & 22-Apr-16 & 501.0394 & 10x4x79 & 545.7981 & 1.204 & 6699 & ND & $ 14.12 \pm 161.95 $ & $ -21.76 \pm 162.19 $ \\
                  &         & 19-May-16 & 527.9511 & 10x4x79 & 544.3413 & 1.205 & 7011 & ND & $ 13.27 \pm 158.54 $ & $ -3 \pm 158.34 $ \\
          \hline
HD\,181569        & 3437940 & 17-Jun-16 & 557.1169 & 4x4x206 & 528.8988 & 1.194 & 6036 & ND & $ -28.0 \pm 338.38 $ & $ 0.73 \pm 338.77 $  \\
                  &         & 24-Jun-16 & 564.0810 & 8x4x206 & 522.6965 & 1.195 & 11972 & ND & $ -3.56 \pm 124.79 $ & $ -0.24 \pm 124.94 $  \\
          \hline
HD\,183280        & 10664975 & 11-Jun-16 & 551.0712 & 1x4x1728 & 526.2097 & 1.199 & 5597 & ND & $ -65.0 \pm 120.8 $ & $ -67.1 \pm 120.4 $ \\
                  &          & 19-Jun-16 & 558.9841 & 1x4x1728 & 524.8147 & 1.199 & 5170 & ND & $ 228.1 \pm 113.9 $ & $ -86.8 \pm 113.8 $ \\
          \hline
HD\,185115        & 9775454  & 14-May-16 & 523.1233 & 1x4x519 & 523.6516 & 1.195 & 3671 & ND & $-15.5 \pm 24.7$ & $26.7 \pm 24.8$ \\
                  &          & 17-May-16 & 526.1238 & 1x4x519 & 523.1970 & 1.195 & 3676 & ND & $0.1 \pm 20.4$ & $10.4 \pm 20.4$ \\
                  &          & 18-Jun-16 & 558.1102 & 1x4x519 & 523.1141 & 1.195 & 3601 & ND & $-18.6 \pm 22.5$ & $25.4 \pm 22.4$ \\
          \hline
HD\,187254        & 8703413 & 19-Jun-16 & 559.0547 & 1x4x1145 & 531.3430 & 1.196 & 1968 & ND & $ 0.8 \pm 6.3 $ & $ -4.6 \pm 6.3 $ \\
          \hline
HD\,226284        & 5473171 & 9-Jun-16  & 549.0866 & 5x4x285 & 520.7524 & 1.191 & 11400 & ND & $ 110.14 \pm 244.56 $ & $ 170.04 \pm 244.96 $ \\
                  &         & 16-Jun-16 & 556.0860 & 5x4x285 & 522.4294 & 1.190 & 9702 & ND & $ -184.28 \pm 355.0 $   & $ 75.38 \pm 355.22 $ \\
                  &         & 18-Jun-16 & 558.0266 & 9x4x285 & 522.4064 & 1.190 & 12722 & ND & $ -53.37 \pm 412.01 $ & $ -133.76 \pm 412.01 $ \\
\hline
\hline
\end{tabular}
\caption{Table of key parameter values for the averaged nightly profiles for each of the 12 targets considered in this study. Columns 3 and 4 display the dates the observations were taken, in Gregorian and Mid-Heliocentric Julian Dates respectively. Column 5 shows the exposure time and observing strategy utilised for each target, while columns 6 and 7 provide the mean wavelength and mean Landé factor respectively for each observation. Column 8 provides the mean SNR for the LSD I profiles. The results of the FAP algorithm are shown in column 9, and the longitudinal field measurements taken from the Stokes V ($B_l$) and N ($N_l$) profiles are provided in columns 10 and 11 respectively.}
\label{tab:full_table}
\end{table*}

For this LSD calculation, we used a mask from the Vienna Atomic Line Database \citep[VALD3,][]{piskunov1995,kupka1999,ryabchikova2015} as a template, selected with respect to the individual effective temperature and surface gravity of each target, which we then fine-tuned by removing hydrogen lines, telluric absorption features, or any other features that would have detrimental impacts on the quality of the spectrum. By comparing the resulting mask to the observed spectrum, we then adjusted the line depths to best represent the star, following the procedure described in \cite{grunhut2017}. 

\begin{figure*}
	\includegraphics[width=17cm]{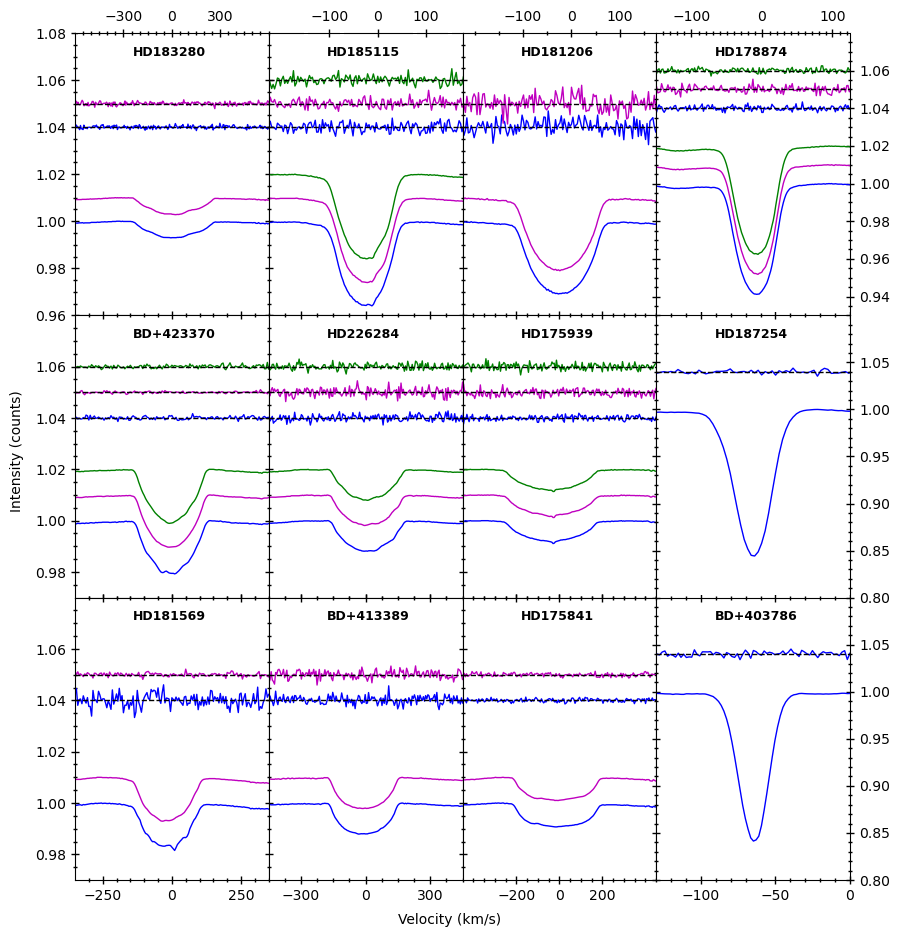}
    \caption{Stokes I (lower) and V (upper) profiles for each target star. When they exist, each subsequent night of observation for a particular target is displayed in blue, magenta and green, respectively. Variations in x- or y-ranges along the top and right-hand side apply \emph{only} to the plots immediately adjacent. The Stokes V profiles have been magnified by a factor 30 to improve legibility, and $V/I_c=0$ is shown by a dashed black line.}
    \label{fig:stokes_profiles}
\end{figure*}

To improve the SNR further and reduce the impact of potential remnant spurious signals, we performed a weighted average on the observations for each night, in order to generate a mean profile per night which would be used for the remainder of the analysis process. Additionally, a simple linear fitting was performed on the Stokes I profiles, in order to better normalise the LSD profile and smooth out any offset. This normalisation factor was also applied to Stokes V and N. This process will increase accuracy in the magnetic field calculations to come. The mean wavelength and Land\'e factor, as well as the SNR for each average nightly profile, are visible in columns 5, 6, and 7 of Table~\ref{tab:full_table} respectively. The final Stokes I and V profiles are displayed in Fig.~\ref{fig:stokes_profiles}. \revision{The N profiles were flat and showed no signs of pulsations, therefore we do not show them  in the figure.}\\

\section{Results}
\subsection{Magnetic status and longitudinal field}
We calculated the longitudinal field values $B_l$ for the averaged nightly profiles. The values are available in column 10 of Table~\ref{tab:full_table}, along with the values for $N_l$ in column 11, which are calculated from the N profiles by applying the same methods used for calculating $B_l$ from the Stokes V profiles. It is quickly apparent that, taking into account their errors, the values for $B_l$ and $N_l$ are consistent with zero within 2$\sigma$ in all cases.

However, $B_l$ can be zero even if the star is magnetic, e.g. if the magnetic field is dipolar and seen edge-on. A Stokes V signature is then present but it is symmetrical around the center of the line, leading to $B_l$=0. Therefore we applied a False Alarm Probability (FAP) algorithm to the averaged nightly profiles for each target, to check for the presence of a magnetic signature. This resulted in no detections in V nor in N for any target. For a definite detection the algorithm requires a value of $\text{FAP} \lesssim 10^{-5}$, and a value of $10^{-5} \lesssim \text{FAP} \lesssim 10^{-3}$ for a marginal detection \revision{\citep{donati1992}}. A value of $\text{FAP} \gtrsim 10^{-3}$ is determined to be a non-detection, as was the case for all our targets.

\subsection{Upper limit on undetected fields}

\begin{figure*}
	\includegraphics[width=17cm]{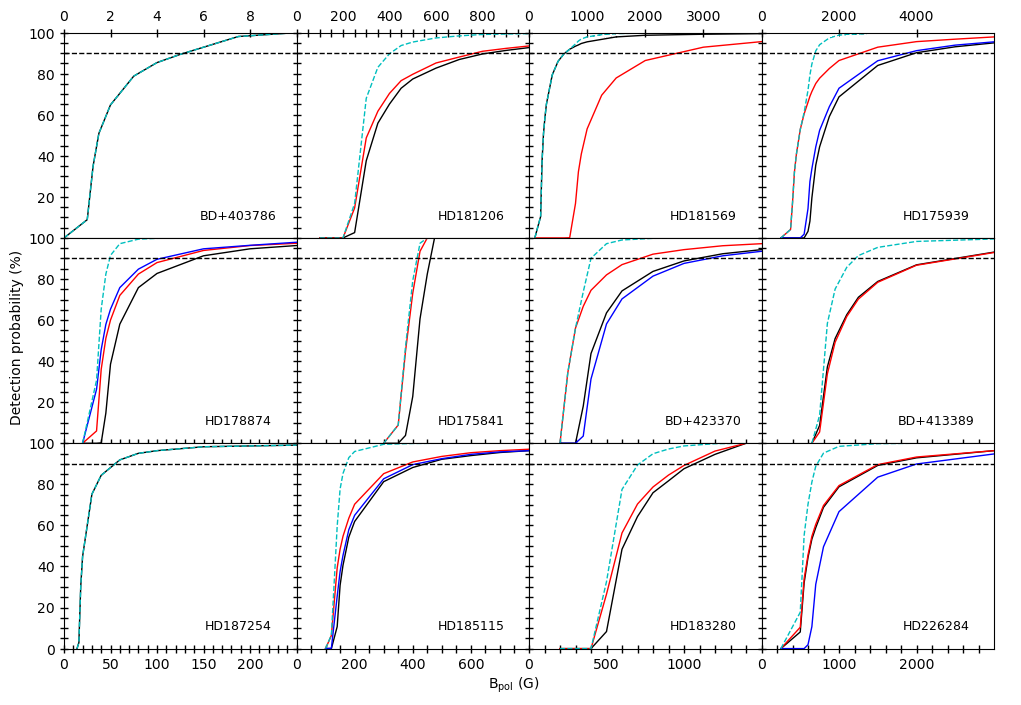}
    \caption{Detection probability for the observations of each star as a function of the magnetic polar field strength. The curves represent the first (black), second (red), third (blue) and combined (cyan dashed) profiles. The horizontal black dashed line represents a 90 per cent detection probability. Plots in the same column share an x-axis range, unless stated otherwise above.}
    \label{fig:upperlims}
\end{figure*}

As a result of the lack of magnetic detections for all of the targets, we performed modelling in order to determine the maximum dipolar field strength ($B_{\rm pol}$) values of a magnetic field that might have remained hidden in the noise of the data. This would allow us to provide an upper limit to $B_{\rm pol}$ for each target, and perhaps conclude on whether these stars are still magnetic candidates. This algorithm calculates 1000 oblique dipole models for each of the LSD Stokes V profiles using random values for inclination angle $i$, obliquity angle $\beta$, rotational phase, and a white Gaussian noise with a null average and a variance corresponding to the SNR of each profile. For a more detailed description of the process, we refer the reader to \citet{neiner2015b}. We then calculated the rate of detections from these 1000 models, and plotted the results in Fig.~\ref{fig:upperlims}.

\begin{table}
\centering
\begin{tabular}{c@{\,}cc@{\,}c@{\,}c@{\,}c}
\hline
Target ID & B$_{\text{pol,max}}$ & $\widetilde{B}_{\text{pol,max}}$ & T$_{\text{eff}}$ & $v$sin$i$ & L  \\
KIC ID     & (G)           & (G)      & (K)              & (km\,s$^{-1}$) & (L$_{\odot}$)  \\
\hline \hline
BD\,+40\degree3786$^{a,e}$ & 5         & 5        & 6800 $\pm$ 250   & 15 $\pm$ 3 & 148 $\pm$ 15    \\
5965837 & & & & & \\
\hline
BD\,+41\degree3389$^{a,b,e}$ & 2500      & 1213      & 8107 $\pm$ 70    & 149.7 $\pm$ 7 & 31 $\pm$ 2  \\
6289468                    & 2532      &          &                  &               &  \\
\hline
BD\,+42\degree3370$^{a,e}$ & 718       & 405      & 7203 $\pm$ 250   & 121 $\pm$ 3 & 35 $\pm$ 4  \\
6951642               & 1086      &          &                  &             &    \\
                        & 1162       &          &                  &             &    \\
\hline
HD\,175841$^{a,c,e}$  & 421       & 415      & 8400 $\pm$ 150   & 191.0 $\pm$ 15 & 60 $\pm$ 2  \\
4989900           & 461       &          &                  &                &     \\
\hline
HD\,175939$^{b,e}$  & 2554      & 1383     & 7891 $\pm$ 62    & 192.8 $\pm$ 11.8 & 34 $\pm$ 2  \\
6756386       & 3756      &          &                  &                  & \\
                & 4041      &          &                  &                  & \\
\hline
HD\,178874$^{b,e}$  & 106        & 49       & 6248 $\pm$ 57    & 39.7 $\pm$ 1.2 & 19 $\pm$ 2  \\
11498638       & 117        &          &                  &                &  \\
                 & 143        &          &                  &                &  \\
\hline          
HD\,181206$^{a,b,e}$  & 770       & 406      & 7478 $\pm$ 41    & 85.1 $\pm$ 2.5 & 8 $\pm$ 1   \\
9764965           & 830       &          &                  &                &  \\
\hline          
HD\,181569$^{a,d}$  & 622       & 622      & 7700 $\pm$ 120   & 120 $\pm$ 5 & 54 $\pm$ 5  \\
3437940           & 2554      &          &                  &             &    \\
\hline          
HD\,183280$^{a,e}$  & 1024      & 713      & 7892 $\pm$ 250   & 243 $\pm$ 10 & 93 $\pm$ 9 \\
10664975          & 1069      &          &                  &              &   \\
\hline
HD\,185115$^{a,d}$  & 385       & 172      & 7050 $\pm$ 150   & 70 $\pm$ 5 & 10 $\pm$ 2  \\
9775454           & 420       &          &                  &            &     \\
                    & 445       &          &                  &            &     \\
\hline          
HD\,187254$^e$      & 55        & 55       & 8000 $\pm$ 150   & 15 $\pm$ 2 & 5 $\pm$ 2   \\
8703413 & & & & & \\
\hline
HD\,226284$^{a,e}$  & 1580      & 719      & 7784 $\pm$ 250   & 164 $\pm$ 5 & 48 $\pm$ 5  \\
5473171       & 1617      &          &                  &             &    \\
                & 2077      &          &                  &             &    \\
\hline
\end{tabular}
\caption{Upper dipolar field strength limit (in G) for both the individual (column 2) and combined (column 3) profiles of each target. The values of effective temperature and luminosity used in the HR diagram are displayed in columns 4 and 6, respectively. Rotational velocity values are provided in column 5. Superscript indices indicate the origin of the displayed stellar parameters: a) \citet{lampens2018}, b) \citet{tkachenko2012}, c) \citet{tkachenko2013}, d) \citet{catanzaro2011}, e) \citet{catanzaro2014}.}
\label{tab:upperlims}
\end{table}

Thanks to the fact that in many cases we have multiple observations, we can combine the calculated upper limit statistics in order to extract a more strict value, taking into account that a magnetic field has not been detected in any of the observations. We can achieve this using the following equation:

\begin{equation}
    P_{\text{comb}} = 100 \left[1-\prod^n_{i=1} \frac{(100-P_i)}{100}\right]
\end{equation}

where $P_i$ is the detection probability for the i$^{th}$ observation and P$_{\text{comb}}$ is the detection probability for n observations combined. All probabilities are expressed in percentage. 

As an example, if for a particular target we have three observations available with probabilities of 70, 80, and 90 per cent respectively that no field stronger than 1000 G was detected, the combined probability with the same condition is 99.4 per cent. 

The results of both the initial upper limit calculations, assuming a 90 per cent detection probability, for each individual profile as well as those using the combined power equation above are detailed in Table~\ref{tab:upperlims}. The upper limit of $B_{\rm pol}$ values vary between 5 and $\sim$1400 G depending on the star. 

\section{Discussion}
\subsection{General Results}
None of the targets observed displayed any significant magnetic signal. Seven of our twelve targets have $T_{\rm eff} \gtrsim 7500$ K. Previous research indicates that about 10 per cent of such stars host a strong magnetic field \citep{grunhut2015,grunhut2017,wade2013}. Statistically, we should expect to see one of these stars display a magnetic signature. Moreover, our targets were not randomly selected stars, but candidates that were pre-selected as good magnetic candidates from a wider sample through rotational modulation signatures. With this kind of pre-selection, \cite{buysschaert2018} obtained a magnetic detection rate of $\sim$70 per cent in late-B to mid-A-type stars (which displayed relatively simple lightcurves dominated by rotation with some showing additional low-amplitude signals possibly due to pulsation), and we should then expect $\sim$5 magnetic detections in our sample of seven stars with $T_{\rm eff} \ge 7500$ K. The non-detection of a field in any of our targets is thus a significant null result. 

Moreover, magnetic OBA stars typically have a magnetic field strength of the order of 3 kG, though it can range between 300 G and 30 kG \citep[e.g.][]{shultz2019}. From the results of the upper limits modelling for the seven targets with $T_{\rm eff} \ge 7500$ K, we can see that all seven stars have upper limits below 3 kG and six of the seven stars have upper limits around 700 G or smaller. If the magnetic field of (hotter) $\delta$\,Sct stars were the fossil fields observed in 10 per cent of OBA stars, we should have detected them in some of our targets. 

Therefore, either our sample displays a frequency of magnetic field representation lower than the average 10 per cent of OBA stars or, if a magnetic field is present in these stars, its strength is well below the average field for OBA stars. 

Five of our twelve targets have $T_{\rm eff} \lesssim 7500$ K. The three discoveries of magnetic $\delta$\,Sct stars with $T_{\rm eff} \lesssim 7500$ K so far have fields that are not strongly dipolar but instead are complex fossil, dynamo, or ultra-weak fields with strength between $\sim$1 and $\sim$1000 G. For our four cooler targets (BD+40\degree3786, BD+42\degree3370, HD\,178874, HD\,185115), our respective upper limits of 5, 405, 49, 172 G (Table \ref{tab:upperlims}) are quite good, especially for BD+40\degree3786 and HD\,178874. Statistics is difficult on only four targets, but the fact that we did not detect any field while these four targets were carefully pre-selected as magnetic candidates is still intriguing. 

In the following we investigate why the targets in our sample may not be magnetic. 

\subsection{Stellar evolution}

A possible factor that we considered is what impact the evolutionary stage of these stars might have on their potential magnetic field strength. Indeed, in the case of fossil fields, considering magnetic flux conservation, older, more evolved stars must have weaker fields at their surface than their younger counterparts \citep{neiner2017}. Moreover, \cite{bagnulo2006,landstreet2007,landstreet2008,fossati2016, shultz2019} showed that an additional intrinsic decay of the field strength may occur. For cooler stars as well, more evolved stars have typically weaker magnetic field strengths \citep[e.g.][]{vidotto2014} than their progenitors.

\begin{figure*}
	\includegraphics[width=12cm]{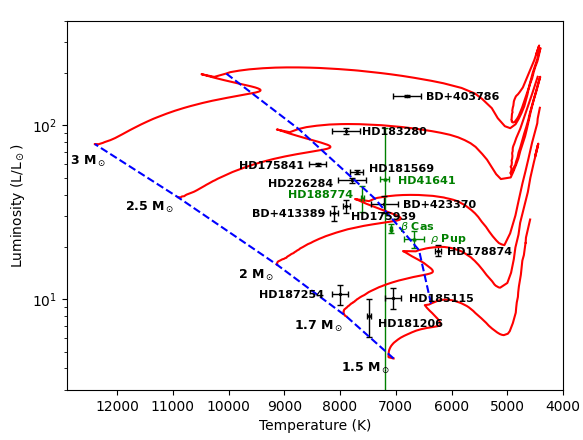}
    \caption{Evolutionary tracks of model stars with Sun-like metallicity and rotation, and masses of 1.5, 1.7, 2.0, 2.5 and 3 $M_\odot$ respectively. Blue dashed lines correspond to ZAMS and TAMS. Black dots represent this sample and green dots represent confirmed magnetic $\delta$\,Sct stars, with datapoints labelled accordingly.}
    \label{fig:evo_tracks}
\end{figure*}

To test this factor, we generated a set of evolutionary tracks for stars with Sun-like metallicity ($Z=0.014$) and rotation ($V/V_c=0.4$), shown in Fig.~\ref{fig:evo_tracks}, using models from the Université de Genève database and detailed in \citet{ekstrom2012}, over which we plotted the stars from our sample. The values for effective temperature $T_{\text{eff}}$ and luminosity $L$ for our targets and their respective errors were taken from an array of surveys \citep{latham2005, tkachenko2012, tkachenko2013, catanzaro2011, catanzaro2014, lampens2018} and are displayed in columns 4 and 5 of Table~\ref{tab:upperlims}. Where unavailable, we determined them using the following equation, valid in the intervals $3.2 < \log g < 4.7 \; ; \; 3.690 < \log T_{\text{eff}} < 3.934$, as determined by \citet{catanzaro2014}:

\begin{equation}
\begin{split}
    \log L/L_\odot = (-15.46 \pm 0.34) &+(5.185 \pm 0.08) \log T_{\text{eff}} \\
    &- (0.913 \pm 0.014) \log g
\end{split}
\end{equation}

To more easily identify the evolutionary stage, the blue dashed lines visible in Fig.~\ref{fig:evo_tracks} were generated, representing the Zero Age Main Sequence (ZAMS) and Terminal Age Main Sequence (TAMS). The ZAMS was determined from the initial values of the models, which corresponded well with a classical mass-luminosity relation for the given parameters, and the TAMS was set to determine the time at which the mass fraction of hydrogen in the core was lower than $X_c (H) =10^{-5}$.

Over the next few subsections, we detail the results for the various stars in the sample, ordered by evolutionary stage, and draw individual conclusions.

\subsubsection{Main sequence stars}
In Fig.~\ref{fig:evo_tracks}, the following stars are squarely located within the main sequence (MS). We consider BD+41\degree3389, HD\,175841, HD\,175939, HD\,181206, HD\,187254, and HD\,185115. The LSD+FAP method provided no magnetic field detection for these targets, and we determined maximum field strengths at 90 per cent probability of 1213, 415, 1383, 406, 55, and 172 G respectively from the upper limits algorithm. Except for HD\,185115, the other five stars have $T_{\rm eff} \sim 7500$ K or more. A typical fossil field is thus expected in a few of those targets but not detected, in spite of the very good upper limit for at least three targets. For HD\,185115, the upper limit of 172 G is probably not sufficient if the field of this cooler star is a dynamo field. 

\subsubsection{Terminal age main sequence stars}
We consider here stars that are located at the end of the MS or the very beginning of the red giant branch (RGB) in Fig.~\ref{fig:evo_tracks}. Here we consider BD+42\degree3370, HD\,181569, and HD\,226284, with respective maximum field strengths with 90 per cent probability of 405, 622, and 719 G, from the upper limits algorithm. Once again, there were no detections for any of these stars through the LSD+FAP methods, although the upper limit values around 600-700 G for the two hotter stars (HD\,181569 and HD\,226284) are quite good with respect to the expected fossil field strength.

For the cooler star BD+42\degree3370, \citet{samadi2022} performed a specific analysis of the \emph{Kepler} lightcurve  with the goal of determining the nature of the dominant frequencies, such as those presented in Fig.~\ref{fig:periodograms}. They used the \emph{Kepler} and \emph{TESS} lightcurves to measure the magnetic activity index S$_{\rm ph}$ (adopting a stellar rotation frequency of 0.721 c/d), and found S$_{\rm ph}$ to vary with a cycle of length $\sim3.2$ yr, which they attributed to a possible signature of long-term stellar activity, and discussed their results in light of the presented spectropolarimetric measurements. Despite the fact that this star does not present Zeeman signatures that might indicate the presence of a magnetic field, and taking into account the upper limits value we acquired in this study, they could not rule out the possibility of a magnetic field being responsible for some of the frequencies observed arguing that, because the star had been observed spectrophotometrically in a state of low activity, the magnetic field could be under-represented. In any case, our upper limit of 405 G for this star is probably not sufficient to detect a dynamo field. 

\subsubsection{Red giant branch stars}
Finally, some stars are evolved, progressing along the RGB. We consider BD+40\degree3786, HD\,178874, and HD\,183280 to be at this stage of their life cycles, and we found values of 5, 49, and 713 G (Table \ref{tab:upperlims}) respectively from the upper limits algorithm, and no magnetic field detections through the LSD+FAP method for these stars. For BD+40\degree3786, our excellent upper limit allows us to conclude that the magnetic field is either ultra-weak (as in the case of $\rho$\,Pup) or nonexistent. For HD\,178874 a weak dynamo field, similar to the one of $\beta$\,Cas, could have remained hidden in this evolved cooler star. On the other hand, the upper limit for HD\,183280 is quite high for an evolved hot star, it is therefore not surprising that we did not detect a field in this target. However, the value found for its rotational velocity ($v$sin$i$ $\approx 243$ km\,s$^{-1}$) is unusually high for a star on the RGB, therefore this star might be less evolved than found here and the magnetic field should have been detected if the star is on the MS.

\subsubsection{Evolution study summary}
From this evolution study, we thus conclude that we can explain the absence of detection of a magnetic field in at least two stars (BD\,+42\degree3370, HD\,178874), and possibly in HD\,183280, but not in the other targets. 

\subsection{Re-examining {\it Kepler} data in light of the magnetic non-detections}
\subsubsection{Case-by-case discussion}
\label{sec:case_by_case}
Considering the results of magnetic non-detections for this sample, it is worthwhile to re-examine the {\it Kepler} data. This is briefly done here in the same order as in Fig.~\ref{fig:periodograms}. We recall that the presence of a set of harmonic frequencies points to rotation and thus to a magnetic candidate. When two sets of harmonics are present, it may suggest rotation from two different stars (in a binary or multiple system).

{\it BD+40\degree3786 = KIC 5965837:} No harmonic series were found. The {\it Kepler} signals seem consistent with $\delta$\,Sct pulsation, but with relatively low frequencies. The strongest signals are at 4.05 d$^{-1}$ and 3.39 d$^{-1}$. This star was determined by \citet{lampens2018} to be a $\rho$\,Pup star. In this case it may host an ultra-weak field. When comparing the magnetic analysis of $\rho$\,Pup itself with our upper-limit calculation for BD+40~3786 this seems consistent, though such an ultra-weak field could have remained undetected even with the very low upper field limit we achieved here.

{\it BD+41\degree3389 = KIC 6289468:} There are two sets of harmonics, one with a fundamental frequency of 1.6290 d$^{-1}$ and the other at 1.551 d$^{-1}$, the former being significantly stronger. There are many additional low-frequency signals unrelated to the two harmonic series, and which also do not resemble typical $\gamma$\,Dor pulsations. For instance, there is a `comb' of frequencies centred near 1 d$^{-1}$ which are split almost, but not exactly, evenly by $\sim$0.09 -- 0.1 d$^{-1}$. There are $\delta$\,Sct frequencies out to $\sim$31 d$^{-1}$. 

{\it BD+42\degree3370 = KIC 6951642:} There are two sets of harmonics, one with a fundamental frequency of 0.7288 d$^{-1}$ and the other at 0.8008 d$^{-1}$. However, the peak of the former is broad and it is unclear if it forms a genuine harmonic series. There are two wide and richly-populated groups of $\delta$\,Sct frequencies centred near 15 d$^{-1}$ (stronger) and 30 d$^{-1}$ (weaker). This star also has at least one combination frequency, where a low frequency signal (2.08 d$^{-1}$) is equal to the difference between two higher frequency modes (16.44 d$^{-1}$ and 13.48 d$^{-1}$). Another long-period binary \citep{lampens2018}, the frequency analysis of BD+42\degree3370 likely exhibits frequencies originating from both stars. Indeed, we observe two distinct sets of harmonics in the periodograms of Fig.~\ref{fig:periodograms}. Therefore, while the rotational variation hints at dynamo field-like features, it is not guaranteed the field and pulsations are on the same target. 

{\it HD\,175841 = KIC 4989900:} There are no harmonic pairs, and thus no candidate rotation frequency was identified. The most prominent low-frequency feature is a group of signals centred near 2.2 d$^{-1}$ which could be consistent with $\gamma$\,Dor pulsation. The strongest $\delta$\,Sct signals are at 6.17 d$^{-1}$, 8.05 d$^{-1}$, and 8.21 d$^{-1}$, but signals extend out to $\sim$15 d$^{-1}$. Without clear rotational variation, this star may indeed not be magnetic. 

{\it HD\,175939 = KIC 6756386:} There are two sets of harmonics. The higher amplitude sequence has a fundamental frequency of 1.7655 d$^{-1}$, and the other 1.620 d$^{-1}$, with both having a first and second harmonic. The frequency spectrum is highly populated, with signals out to 50 d$^{-1}$. There is a group of at least four frequencies centred near 5 d$^{-1}$ with near-equal spacing of $\sim$0.15 to 0.17 d$^{-1}$. 
HD\,175939 has been determined to be a long-period binary \citep{lampens2018}, and we do observe two sets of harmonics in the frequency analysis, so it is difficult to determine which frequencies originate from which star. Nevertheless, the presence of these sets of harmonics points towards rotational modulation.

{\it HD\,178874 = KIC 11498538:} The only feature in the periodogram is a harmonic series with a fundamental frequency of 0.29982 d$^{-1}$, extending to the third harmonic (but the fundamental and first harmonic have significantly higher amplitude). The SC data reveal rapid variations with a timescale of $\sim$1 hour. However, these do not appear as peaks in the periodogram and thus are not coherent $\delta$\,Sct modes. These may be solar-like oscillations but which are not easily measurable due to the short baseline of the SC data and/or too high of a noise floor. Between rotational variation found in its frequency analysis and cooler surface temperatures ensuring a convective envelope, HD\,178874 presents all the features of a dynamo field. However, our 49 G upper limit value is probably not low enough to detect it.

{\it HD\,181206 = KIC 9764965:} There is one harmonic series with the fundamental frequency at 2.051 d$^{-1}$. There are a few non-harmonic low frequencies. There are not many $\delta$\,Sct frequencies, but compared to the rest of the sample they are relatively high. The temperature of this star puts it at the limit between dynamo and fossil field types. The upper limit value of 406 G is too high to detect a dynamo field, but would be sufficient to detect a typical fossil field. However, we remark that HD\,181206 has been classified as a metal-lined star (with spectral type Am(p)) by \cite{bertaud1960}. All Am stars studied with sufficiently deep spectropolarimetric measurements show a magnetic field but with a typical strength of only a few Gauss \citep{blazere2016}. The strongest field has been detected in the Am star Alhena with ~30 G \citep{blazere2020}. Our observations would not allow to detect such a weak or ultra-weak field.

{\it HD\,181569 = KIC 3437940:} The best candidate rotation frequency is at 0.8659~d$^{-1}$, with a first harmonic and a signal close to, but not exactly at, its second harmonic. There are multiple peaks with amplitudes of $\sim$200 ppm which bear no relation to the 0.8659~d$^{-1}$ signal. These signals do not obviously resemble the typical patterns of $\gamma$\,Dor pulsation, and thus the type of pulsation they represent ($\gamma$\,Dor vs. $\delta$\,Sct) is unclear. Signals are found out to $\sim$19~d$^{-1}$, with the strongest modes at 10.48~d$^{-1}$, 10.33~d$^{-1}$, and 12.02~d$^{-1}$.

{\it HD\,183280 = KIC 10664975:} There are two sets of harmonics. The stronger has a fundamental frequency of 1.2602 d$^{-1}$ and the series extends until at least the sixth harmonic (perhaps suggesting a blended eclipsing binary is responsible). The weaker fundamental frequency of 1.1892 d$^{-1}$ has only a first harmonic. There are low frequency signals unrelated to these two series. At least one of these, at 2.5735 d$^{-1}$ corresponds to the difference between higher frequency $\delta$\,Sct modes (19.8726 d$^{-1}$ $-$ 17.2992 d$^{-1}$, and 21.132 d$^{-1}$ $-$ 18.5590 d$^{-1}$), suggesting mode coupling within the star. 

{\it HD\,185115 = KIC 9775454:} There do not seem to be any harmonic relations among the many low-frequency signals. However, there are two wide frequency groups centred near 1.5 d$^{-1}$ and 3 d$^{-1}$, which could be consistent with $\gamma$\,Dor pulsation. Relatively higher amplitude peaks are evident at 4.1602 d$^{-1}$ and 4.6105 d$^{-1}$, and a signal is seen at their difference (0.450 d$^{-1}$), suggesting some degree of mode coupling in the star. HD\,185115 is the third long-period binary found in this sample \citep{lampens2018,lampens2021}, with an orbital period of 1707 days, though this one does not present rotational variation observed in the previous two. It is thus unlikely to be magnetic. 

{\it HD\,187254 = KIC 8703413:} The only feature in the periodogram is a harmonic series with a fundamental frequency of 0.1464 d$^{-1}$, extending to the third harmonic (but the fundamental and first harmonic have significantly higher amplitude). We see no sign of pulsation. The SC data shows a flare lasting for $\sim$12 minutes and an amplitude of $\sim$600 ppm. Flares were not seen in any other star in the sample, but given their short duration their detection is only possible with SC data. 
Flares are often good indicators of dynamo fields, however with a surface temperature around 8000 K it is unlikely that this is the case here. Instead, it is more likely that the flare originates from an unseen companion, as described for similar stars by \citet{pedersen2017}. 
HD\,187254 is an Am star with a metal-rich surface composition ([Fe/H]=0.254) and according to \citet{balona2015b} its {\it Kepler} lightcurve shows travelling features that could be caused by spots. Like for HD\,181206, only weak and ultra-weak fields have been measured in Am stars, so our low upper
limit value of 55 G might not be sufficient to detect such a field.

{\it HD\,226284 = KIC 5473171:} There are two sets of harmonics, one with a fundamental frequency of 1.405 $^{-1}$ (with one harmonics), and the other at 0.971 $^{-1}$ (with two harmonics). Their amplitudes are similar and it is difficult to determine which set is more likely to correspond to rotation of the $\delta$\,Sct star. There are many low-frequency signals unrelated to these harmonic series, which then are likely to be pulsational. However, there are no clear patterns to aid in their identification without a more detailed analysis. $\delta$\,Sct pulsation extends out to $\sim$18 d$^{-1}$, and the dominant signal is at 7.57 d$^{-1}$. There may be a combination frequency between a low-frequency signal at 0.266 d$^{-1}$ and two higher-frequency signals at 14.402 d$^{-1}$, 14.136 d$^{-1}$ (i.e. their difference is the low frequency). Such combinations have been seen in other $\delta$\,Sct stars, and suggests that all of these pulsation frequencies exist in the same star and may be due to non-linear interaction. 

\subsubsection{Results of case-by-case study}
From the above analysis, we conclude that we can explain the absence of detection of a magnetic field in an additional five stars: BD\,+40\degree3786, HD\,175841 and HD\,185115 which show no rotational modulation, HD\,181206 and HD\,187254 which are Am stars.

It should also be noted that of our list of 12 potentially hybrid candidates, only three appear to display frequencies typical of both $\delta$\,Sct and $\gamma$\,Dor variable stars and thus may be genuine hybrids: BD+40\degree3786, HD\,175841 and HD\,185115. Once again, the latter is a confirmed binary, so care should be taken when drawing conclusions regarding the frequencies of this star.

\section{Conclusions}

Although our sample is statistically small and our data likely do not have a sufficient SNR to detect dynamo fields in the four coolest stars of our sample, at least some magnetic stars should have been detected among the eight hottest targets of our sample, considering that we pre-selected them based on the possible existence of a rotational modulation signal. If the chances of magnetic field detection using rotational modulation as a prior is valid \citep[following the work by][]{buysschaert2018}, then the fact that no field was detected with good upper limits amongst the hot stars in our sample is a significant result.

From our case-by-case analysis in the previous section, we can find a plausible explanation for the non-detection of a magnetic field in seven out of twelve targets. Either these stars show no rotational modulation, or our observations lead to a too high upper field limit with respect to the field strength expected from the evolutionary status, chemical peculiarity (Am), or temperature of the star (cool stars with dynamo fields). 

For the remaining few stars (BD\,+41\degree3389, HD\,175939, HD\,181569, HD\,226284, and possibly HD\,183280), we observe rotational modulation in their \emph{Kepler} data and the upper limits we derive are low enough (622 to 1213 G) to detect a fossil magnetic field in these hotter (T$_{\rm eff}$ = 7700-8107 K) targets. All five stars are confirmed $\delta$\,Sct variables from our analysis. 

It is possible that we overestimated the probability of detecting a field in this sample, that the chance of these stars hosting a magnetic field is not more than 10 per cent, and due to the low statistics we were unlucky with the targets chosen for this study. However, if the chances of magnetic field detection using rotational modulation as a prior is valid \citep[following the work by][]{buysschaert2018}, then we should have expected to detect a magnetic field in at least some of these five stars. 

There are two possible explanations for the dearth of magnetic fields in these remaining five stars: 
\begin{itemize}
    \item either the low frequencies observed in the lightcurves of these hybrid candidates are not due to rotational modulation associated to the presence of a magnetic field. They could be related to binarity, e.g. ellipsoidal variation or tidally excited \emph{g-}modes.
    Alternatively, they could be due to the presence of Rossby-modes that are related to the rotational frequency as found in several $\gamma$\,Dor stars \citep{vanreeth2016,takata2020,saio2021}. There could also be a $\gamma$\,Dor companion associated to the $\delta$\,Sct star. However, HD\,175939, HD\,185115 and BD+42\degree3370 are the only three targets of our sample for which (long-period) binarity was found and this explanation does thus not seem adequate for all five stars.
    \item or the magnetic fields of hot $\delta$\,Sct stars are typically weaker (and possibly more complex) than the dipolar fossil fields of OBA stars. This explanation seems plausible considering that the magnetic $\delta$\,Sct stars discovered so far all have fields below $\sim$1 kG (see Table~\ref{tab:delta_scuti}). This could be demonstrated by acquiring deeper spectropolarimetric observations of the five hotter targets listed above or of other bright, low $v$sin$i$, $\delta$\,Sct stars. Since \emph{TESS} observed brighter targets than \emph{Kepler}, including many $\delta$\,Sct stars, good candidates for a search for weak fields in $\delta$\,Sct stars are now available and will be the targets of the MOBSTER collaboration \citep{daviduraz2019, neiner2021}. If weaker fields were confirmed in $\delta$\,Sct stars, they would provide important constraints to stellar models, in particular to the interplay between magnetism, rapid rotation, and pulsations. If these stars are indeed found not to be magnetic even with deeper spectropolarimetric observations, this begs the question of the origin of this discrepancy with normal OBA stars and hotter pulsators \citep{silvester2009}.
\end{itemize}

\section*{Acknowledgements}
We thank the anonymous referee for comments that improved the manuscript. The work herein was based on observations obtained at the Canada-France-Hawaii Telescope (CFHT) which is operated by the National Research Council (NRC) of Canada, the Institut National des Sciences de l'Univers of the Centre National de la Recherche Scientifique (CNRS) of France, and the University of Hawaii. Based on data gathered with NASA’s Discovery mission {\it Kepler} and with the HERMES spectrograph installed at the Mercator telescope, operated on the island of La Palma by the Flemish Community at the Spanish Observatorio del Roque de los Muchachos of the Instituto de Astrof\'{i}sica de Canarias and supported by the Fund for Scientific Research of Flanders (FWO), Belgium, the Research Council of KU Leuven, Belgium, the Fonds National de la Recherche Scientific (F.R.S.– FNRS), Belgium, the Royal Observatory of Belgium, the Observatoire de Genève, Switzerland, and the Thüringer Landessternwarte Tautenburg, Germany. The research leading to these results has (partially) received funding from the KU~Leuven Research Council (grant C16/18/005: PARADISE), from the Research Foundation Flanders (FWO) under the grant agreement G089422N, as well as from the BELgian federal Science Policy Office (BELSPO) through PRODEX grant PLATO. This research has made use of the SIMBAD database operated at CDS, Strasbourg (France), and of NASA's Astrophysics Data System (ADS).

\section*{Data Availability}
The data underlying this article are available in the PolarBase database, at \url{https://polarbase.irap.omp.eu}


\bibliographystyle{mnras}
\bibliography{delscu} 


\clearpage

\begin{appendix}

\section{Rotational signals in HD\,41641 and $\rho$\,Pup} \label{sec:ap1}

\begin{figure}
    \centering
    \includegraphics[width=1\columnwidth]{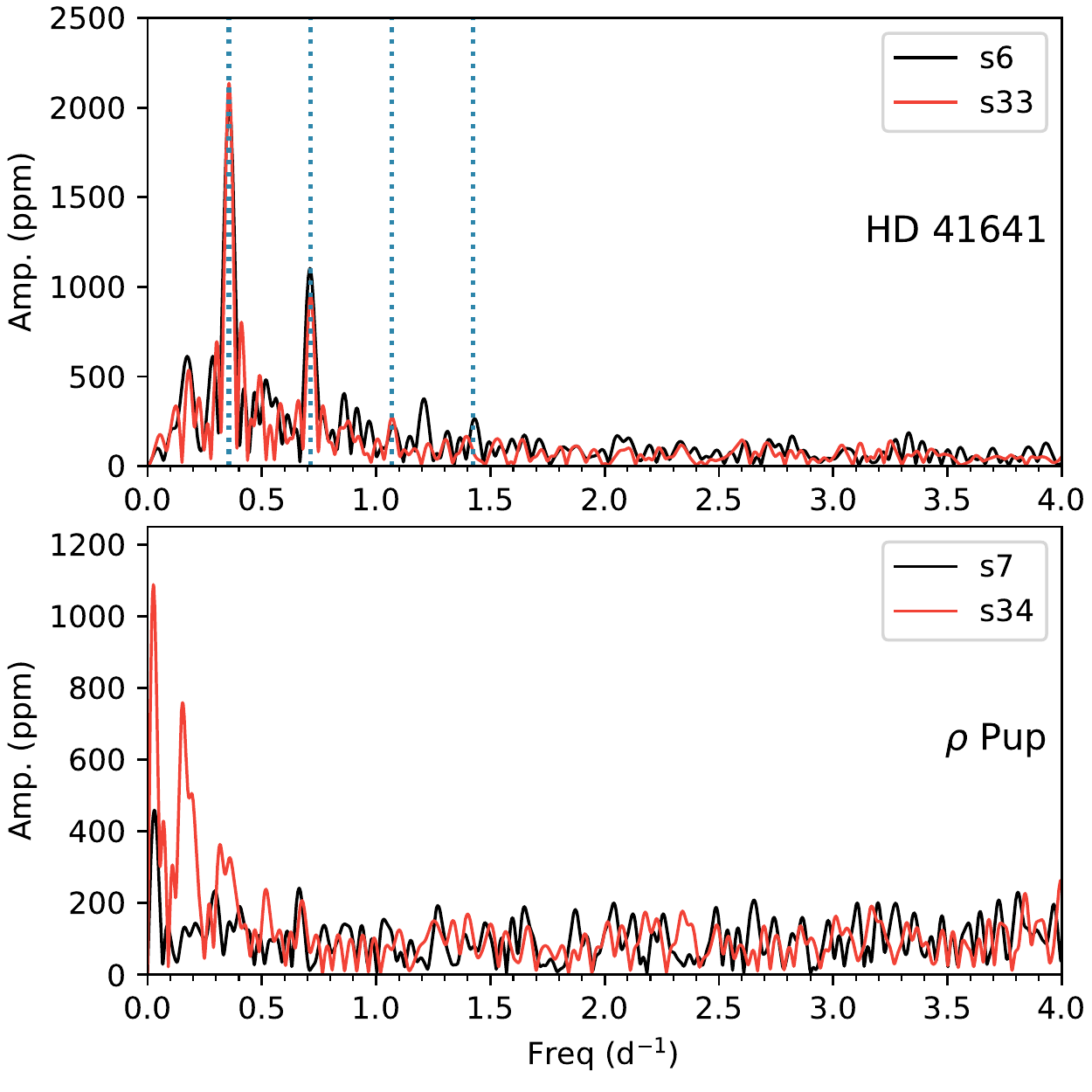} 
    \caption{Frequency spectra from \emph{TESS} for HD\,41641 (top) and $\rho$\,Pup (bottom). For both stars, data from two sectors (each spanning $\sim$27 d) are plotted, with about two years separating the two time strings. For HD\,41641, the rotational frequency and its first three harmonics are indicated with vertical dotted lines. No hint of rotational modulation is seen in the $\rho$ Pup photometry.}
    \label{fig:FTs_appendix}
\end{figure}

High quality space photometry has already been analysed for two of the four known magnetic $\delta$\,Sct stars (HD\,188774 and $\beta$\,Cas; Sec.~\ref{sec:Kepler_analysis}). Space photometry from \emph{CoRoT} was analysed for HD\,41641, but the results were somewhat ambiguous in terms of its rotational period. For $\rho$ Pup, there does not seem to be any published space photometry analysis. We therefore extracted light curves from \emph{TESS} for HD\,41641 and $\rho$ Pup to check for signs of rotational modulation. The low frequency regime is plotted in Fig.~\ref{fig:FTs_appendix} for two (non-consecutive) sectors each for both stars. In HD\,41641, a harmonic sequence consistent with $f_{\rm rot}$ = 0.356 d$^{-1}$ is evident \citep[in agreement with the preferred rotational frequency adopted in][]{thomson2020}. The amplitudes of these signals seem constant over time, suggesting that the surface inhomogeneities are stable in the two years between the two \emph{TESS} sectors, as expected for a fossil field. $\rho$ Pup, however, does not exhibit any sign of rotational modulation. This could be due to a combination of factors, including the low $v$sin$i$ (implying a long rotation period which can be difficult to detect in the short $\sim$27 d \emph{TESS} observing sectors) and the weakness of the field (perhaps causing any surface spots to have a low contrast and thus a small photometric amplitude associated with rotation). 

\section{Rapid non-periodic variations in short cadence data} \label{sec:ap2}

\begin{figure*}
    \centering
    \includegraphics[width=2\columnwidth]{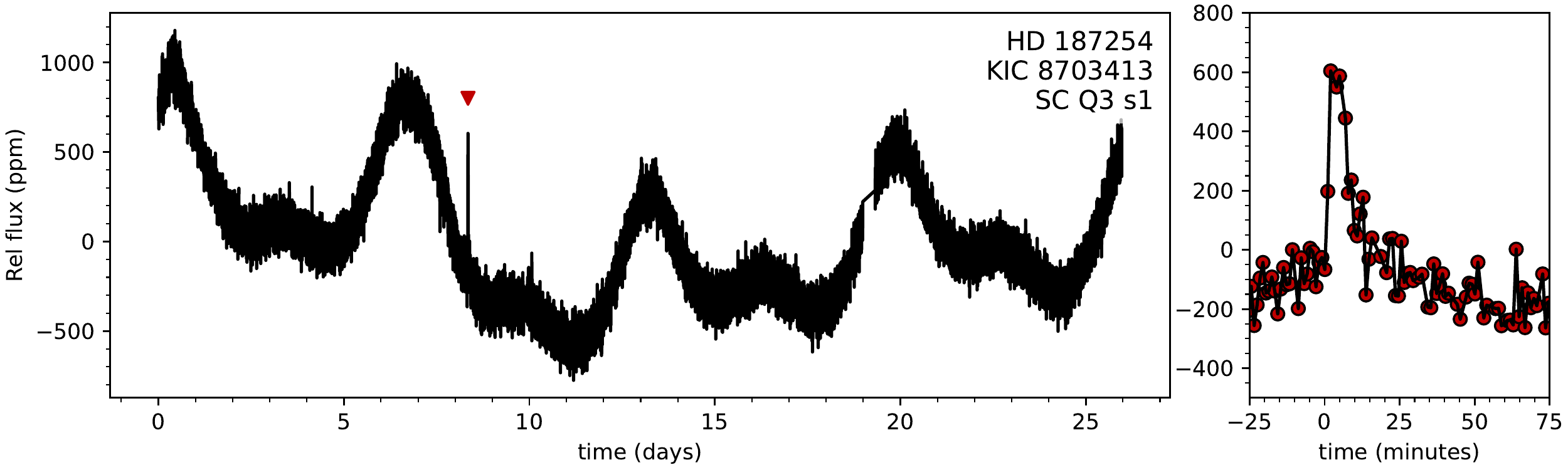} 
    \caption{Short cadence {\it Kepler} light curve for HD\,187254 = KIC 8703413. The left panel shows the entire light curve, with the red triangle indicating the flare. The right panel zooms in on the flare. }
    \label{fig:flare}
\end{figure*}

\begin{figure*}
    \centering
    \includegraphics[width=2\columnwidth]{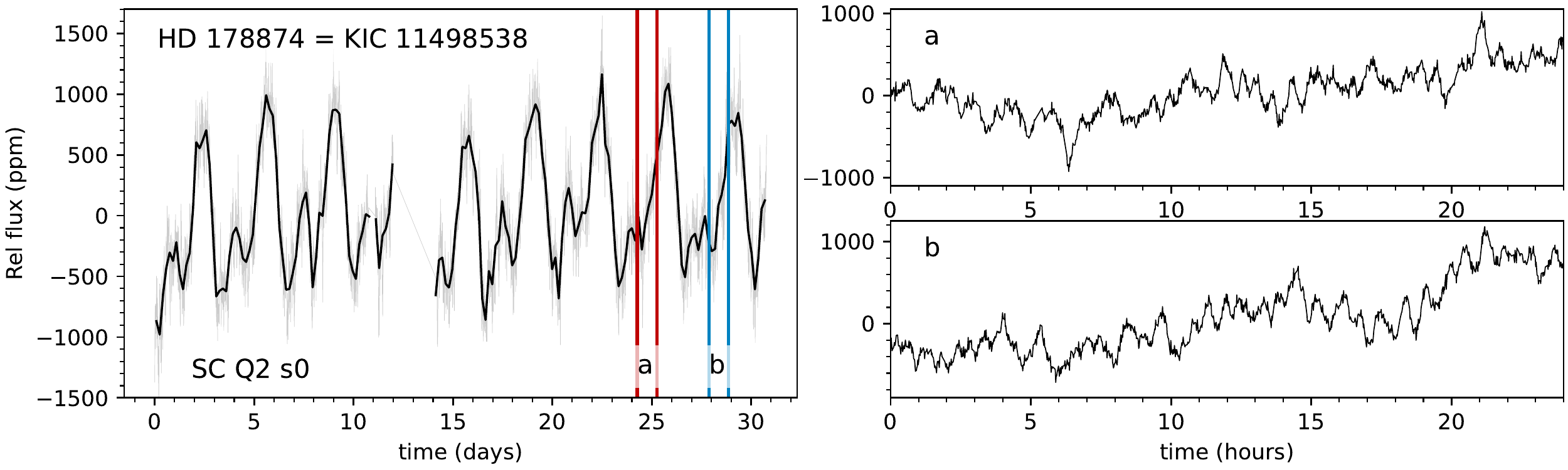} 
    \caption{Short cadence {\it Kepler} light curve for HD\,178874 = KIC 11498538. The left panel shows the entire light curve in lighter grey, with a solid black line plotting the data with four hour bins (to average over the faster variations). Two pairs of vertical lines mark two 1 day sections. The right panels zoom in on these two 24 hour windows, showing the rapid but incoherent oscillations that take place on a timescale of approximately one hour. The selected short observing windows are arbitrary -- similar variations are seen throughout the full time series. }
    \label{fig:fast_osc}
\end{figure*}

The two stars without $\delta$\,Sct pulsation, HD\,187254 (KIC 8703413) and HD\,178874 (KIC 11498538), nonetheless showed their own types of rapid variation. In HD\,187254 (Fig.~\ref{fig:flare}), one flare was observed in the single available short cadence timeseries, which may be an event associated with a lower-mass companion (see Sec.~\ref{sec:case_by_case}). In HD\,178874, fast oscillations are seen throughout the entirety of the short cadence light curves with timescales of roughly one hour (in addition to the much slower rotational variability, Fig.~\ref{fig:fast_osc}). As mentioned in Sec.~\ref{sec:case_by_case}, these features do not appear in the frequency spectrum, nor do they appear coherent in the light curve (right panels of Fig.~\ref{fig:fast_osc}).

\end{appendix}

\bsp	
\label{lastpage}
\end{document}